\documentclass[twocolumn,aps,showpacs,preprintnumbers,amsmath,amssymb]{revtex4}
\usepackage{graphicx}
\usepackage{dcolumn}
\usepackage{bm}
\usepackage{epsfig}
\usepackage{subfigure}

\begin{document}

\bibliographystyle{prsty}

\title{Implications of the Low-Temperature Instability of Dynamical Mean
Theory for Double Exchange Systems}

\author{Chungwei~Lin and Andrew J.~Millis}

\affiliation{ Department of Physics, Columbia University \\
538W 120th St NY, NY 10027}

\begin{abstract}
The single-site dynamical mean field theory approximation to the 
double exchange model is found to exhibit a previously unnoticed 
instability, in which a well-defined ground state which is stable against
small perturbations is found to be unstable to large-amplitude but
purely local fluctuations. The instability  
is shown to arise either from 
phase separation or, in a narrow parameter regime, from the
presence of a competing phase.  The instability is therefore
suggested as a computationally inexpensive means of locating
regimes of parameter space in which phase separation occurs.
\end{abstract}


%%%%%%%%%%%%%%%%%%%%%%%%%%%%%%%%%%%%%%%%%%%%%%%%%%%%%%%%%%%%%%%%%%%%%%%%%%%%%%%%%%%%%%%%%%%%%%%
%%%%%%%%%%%%%%%%%%%%%%%%%%%%%%%%%%%%%%%%%%%%%%%%%%%%%%%%%%%%%%%%%%%%%%%%%%%%%%%%%%%%%%%%%%%%%%%
\maketitle

\section{Introduction}

%(1) DMFT IS IMPORTANT GENERAL TECHNIQUE

Dynamical mean field theory (DMFT) has been widely applied to many strongly
correlated electron systems\cite{Georges}. Since DMFT takes local quantum fluctuations
into account, it is especially successful for models whose many body effect
comes from the on-site interaction, like Hubbard\cite{Hubbard} or Kondo\cite{Kondo}\cite{Chattopadhyay} (double exchange) model.
Correlated systems often exhibit different phases which are quite close in 
energy, and this proximity can lead to phase separation, which is often
important for electronic physics\cite{Dagotto2}. Phase separation is
in principle a "global" property of the phase diagram and requiring
substantial effort to establish:  one
must compute the free energy over a wide parameter range, and then perform a
Maxwell construction.  In this paper we show that within the single-site
DMFT formalism a straightforward calculation at a fixed parameter
value can reveal the presence of phase separation. Specifically, we find
that at zero temperature, the DMFT can give a ground state 
which is stable against small perturbations but is unstable to a large amplitude local
perturbation; at non-zero temperature the standard methods simply fail to converge
to a stable solution. By computing
the free energy and performing a Maxwell construction we show that 
for wide parameter ranges this instability occurs in the regions in which phase
separation exists. In a narrow parameter regime it signals instead the
onset of a different, but apparently uniform, phase.  We therefore propose that 
the instability of the DMFT equations can be used as an approximate, computationally
convenient estimator for the boundaries of the regimes in which phase separation occurs.

The balance of this paper is organized as follows. We first
present the model and then a zero temperature dynamical mean field analysis
explicitly showing the instability.  In section III we calculate the full T=0 phase
diagram in the energy-density plane, establish the regime of phase separation
via the usual Maxwell construction, and extend the treatment to $T>0$. In section IV 
we discuss the implications of this instability. 
Finally in section V we present a brief conclusion.

%%%%%%%%%%%%%%%%%%%%%%%%%%%%%%%%%%%%%%%%%%%%%%%%%%%%%%%%%%%%%%%%%%%%%%%%%%%%%
%%%%%%%%%%%%%%%%%%%%%%%%%%%%%%%%%%%%%%%%%%%%%%%%%%%%%%%%%%%%%%%%%%%%%%%%%%%%%
\section{Double Exchange Model and Dynamical Mean Field Approximation}

\subsection{Double Exchange Model}
In this paper, we consider the single orbital "double exchange"
or Kondo lattice model
of carriers hopping between sites on a lattice and coupled to an array of spins. This model
has been studied by many authors and
contains important aspects of the physics of the "colossal"\cite{Colossal} magnetoresistance
manganites and is also solvable in a variety of approximations, permitting
detailed examination of its behavior. Here we use it to investigate the physical
meaning of a previously unnoticed instability of the dynamical mean field equations.

The model is defined by the Hamiltonian
\begin{eqnarray}
H&=& \sum_{\sigma,i, j} t_{ij}( c^+_{i, \sigma} c_{j,\sigma}+h.c.)
+ \sum_{i,\alpha,\beta} J \vec{S}_i \cdot c^+_{i,\alpha}\,
\vec{\sigma}_{\alpha \beta}\, c_{i,\beta} \nonumber \\ \label{Eqs:H}
\end{eqnarray}
with $i,j$ labeling the sites and the $\vec{S}_i$ denoting the spins.
We assume the spins are classical ($\left[{\vec S}_i,{\vec S}_j\right]=0$)
and are of fixed length. We choose the
convention $|\vec{S}_i|=1$.

The hopping $t_{ij}$ defines an energy dispersion $\varepsilon_k$
and thus a density of states $D(\varepsilon)=\int d^dk/(2\pi)^d \delta(\varepsilon-\varepsilon_k)$.
In our actual computations we specialize to the $d\rightarrow \infty$ limit of the Bethe lattice,
for which $D(\varepsilon)=\sqrt{4 t^2-\epsilon^2}/(2 \pi t^2)$
because the availability of convenient analytical expressions allows us to
accurately compute the small difference between free energies of
different states. We choose energy units such that $t=1$.

We also note that the ground state properties of the model
may be straightforwardly obtained, because for any fixed configuration
of the spins the model is quadratic in the fermions and easily diagonalizable.

%%%%%%%%%%%%%%%%%%%%%%%%%%%%%%%%%%%%%%%%%%%%%%%%%%%%%%%%%%%%%%%%%%%%%%%%%%%%%%
\subsection{Dymanical Mean Field Method}

We now present the dynamical mean field analysis of this model. In
the single site dynamical mean field method\cite{Georges}, one neglects the
momentum dependence of the self energy. The properties of the model
may then be calculated by solving  an auxiliary quantum impurity model, along
with a self consistency condition. The quantum impurity model
corresponding to Eqn(\ref{Eqs:H})  is
specified by the partition function
\begin{eqnarray}
Z_{imp}=\int \,d\vec{S}\, e^{{\cal A}}
\end{eqnarray}
with ${\cal A}=\mbox{Tr}
\log\left[a_{\uparrow}a_{\downarrow}-J \cos\theta
(a_{\uparrow}-a_{\downarrow})-J^2  \right]$
where the trace is over frequency, and $\cos\theta=\hat{z}\cdot\vec{S}$. 
$S$ is determined by a spin-dependent mean field function $a(\omega)$.
In a magnetic phase, $a_{\uparrow}\neq a_{\downarrow}$. Note
that the assumption of classical core spins means that $\int
\,d\vec{S}$ denotes a simple scalar integral over directions of the
core spin $\vec{S}$, and that no Berry phase term
occurs in the argument of the exponential.

The Green function $G_{imp}$ and self energy $\Sigma(\omega)$
of the impurity
model are given by
\begin{eqnarray}
G_{imp, \sigma}(\omega) &\equiv& \frac{\delta \log Z_{imp}}{\delta
a_{\sigma}(\omega)}  \nonumber \\
\Sigma_{\sigma}(\omega) &\equiv& a_{\sigma}-G_{imp, \sigma}^{-1}
\label{Eqs:DMFT1}
\end{eqnarray}
$a$ is fixed by requiring the impurity Green's function $G_{imp}$ equals to the
local Green's function of the lattice problem. The form of the self consistency equation
depends on the state which is studied. For a ferromagnetic (FM) state, it is
\begin{eqnarray}
G_{imp, \sigma}=\sum_{\vec{k}\subset \mbox{BZ}}
(\omega+\mu-\epsilon_{\vec{k}}-\Sigma_{\sigma}(\omega))^{-1} 
\label{Eqs:DMFT2-0}
\end{eqnarray}
while for a 2 sublattice antiferromagnetic (AF) state,
\begin{eqnarray}
G_{imp, \sigma}=
\sum_{\vec{k}\subset \mbox{RBZ}}
\left(
\begin{array} {cc}
\xi_1 &  -\epsilon_{\vec{k}} \\
-\epsilon_{\vec{k}} & \xi_2
\end{array}
\right)^{-1}  
\label{Eqs:DMFT2}
\end{eqnarray}
where $\xi_{1\,(2)}=i \omega_n+\mu-\Sigma_{\uparrow\,(\downarrow)}$ and the $\vec{k}$
-sum is over tje reduced Brillouin zone (RBZ). The two equations become equivalent in the
paramagnetic (PM) state where $\Sigma_{\uparrow} = \Sigma_{\downarrow}$

The solution of  Eqs [\ref{Eqs:DMFT1} to \ref{Eqs:DMFT2}] determines the magnetic phase,
the single particle properties, and the free energy.  In particular, in the dynamical mean field
approximation the Gibb's free energy is  \cite{Georges}\cite{FE}
\begin{eqnarray}
\frac{\Omega}{N}=\Omega_{imp}-T \mbox{Tr}\log
G_{\sigma}(i \omega_n) -T\mbox{Tr}
\log G^{-1}_{\sigma}(i \omega_n, \vec{k}) 
\label{Eqs:LatFreeEnergy}
\end{eqnarray}
where the trace is over the spin and lattice degree of freedom.
The Helmholtz free energy is $F(n)=\Omega+\mu \, N$. 
At zero temperature, the ground state energy (Helmholtz free energy)
is
\begin{eqnarray}
\frac{E}{N}= \mbox{Tr} [ 
G(i \omega_n, \vec{k}) (i\omega_n+\mu) ]
\label{Eqs:T0FreeEnergy}
\end{eqnarray}

We also note that the solution $a_{\sigma}(i \omega_n)$ defines an effective potential for the
core spin, which depends on the angle $\theta$ between the core
spin and local magnetization direction, so that
\begin{eqnarray}
Z_{imp}\rightarrow Z_0 \int \,d\cos\theta\,
e^{-E_{eff}(\cos\theta)/T}
\end{eqnarray}
with
\begin{eqnarray}
E_{eff}(\cos\theta)=-T \sum_{i \omega_n} \log\left[ 1-\frac{J
\cos\theta(a_{\uparrow}-a_{\downarrow}) +J^2
}{a_{\uparrow}a_{\downarrow}}\right]
\label{Eqs:Eeff}
\end{eqnarray}

\subsection{Phase Boundaries and Maxwell Construction}
The model is known to exhibit ferromagnetic, spiral\cite{Spiral} and
commensurate antiferromagnetic phases.
For our subsequent analysis, an accurate determination of phase boundaries
will be important. We therefore present here a few calculational details.

We require the $T=0$ phase boundary separating the ferromagnetic
and spiral phases\cite{Chattopadhyay}\cite{Spiral}. The energy of a spiral state may most easily be found
by performing a site-dependent spin rotation to a basis in which the spin
quantization axis is parallel to the local spin orientation. The problem may
then be easily diagonalized.  For the infinite dimensional Bethe lattice one finds,
for a diagonal spiral of pitch $\phi$, that in the rotated basis, 
the local Green function is given by\cite{Chattopadhyay}
\begin{eqnarray}
\tilde{G}^{-1}_{\sigma}(\omega)  =\omega+\mu-\sigma J
-t^2\cos^2(\frac{\phi}{2}) \tilde{G}_{\sigma} - t^2 \sin^2
(\frac{\phi}{2}) \tilde{G}_{\bar{\sigma}} \nonumber \\ 
\end{eqnarray}
where $\phi$ is the angle between two nearest neighbor
magnetization, and tilde is used for the spiral states. 

To locate the $T=0$ FM/spiral second order phase boundary, 
it suffices to expand ground state energy $E(\mu, \phi)$ 
(Eqn(\ref{Eqs:T0FreeEnergy})) to second order in $\phi$.
The energy difference between FM and spiral states is 
\begin{eqnarray}
dE|_{N} = \frac{\partial \Omega}{\partial \phi} d\phi = \left(\frac{\phi}{2} \right)^2 \, T \sum_n i\omega_n \,dG(i \omega_n, \mu) 
\end{eqnarray}
with $dG$ equaling to
\begin{eqnarray}
d\,G_{\sigma} = \tilde{G}_{\sigma} - G_{\sigma}  = \frac{ t^2
\left( G_{\sigma} - G_{\bar{\sigma}}\right) }{t^2-
G_{\sigma}^{-2}}
\end{eqnarray}
where $\bar{\sigma}=-\sigma$. The FM/spiral phase boundary is determined by 
$dE|_{N}=0$.

%----------------------------------------------------------------------------------
\begin{figure}[ttt]
%\vspace*{-1.6cm}
%\hspace*{-0.5cm}
%\includegraphics[width=\hsize]{Fig111.eps}
   \epsfig{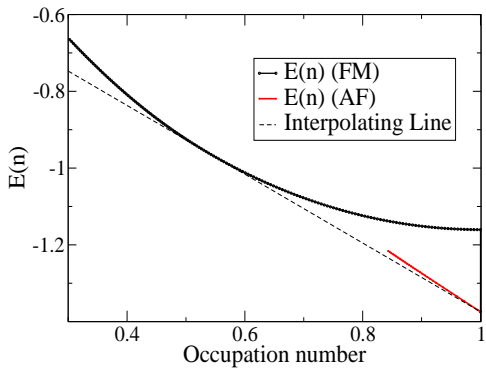}
   \caption{Energy as function of particle density at $J=1$, $T=0$, illustrating 
   Maxwell construction. Heavy solid line: FM state. Light solid line: 2 sublattice 
   AF state. Dashed line: Maxwell construction interpolatiion.  }
   \label{Fig:PST0}
\end{figure}
%--------------------

The model also exhibits phase separation in some regimes.  To determine the
boundaries of the regime where phase separation occurs, we use the DMFT method to compute
the Helmholtz free energy as a function of occupation number
$F(n)$ (Eqn(\ref{Eqs:LatFreeEnergy}))
and then perform the Maxwell construction.
An example is shown in Fig \ref{Fig:PST0}. We find that
in fact that over much of the phase diagram a phase separation
between FM and $n=1$ AF states preempts the formation of
spiral or $n<1$ AF state. The general structure of our
phase diagram agrees with earlier work\cite{Dagotto}\cite{Guinea}, but
the precise locations of phase boundaries differ by roughly $10\%$.
%%%%%%%%%%%%%%%%%%%%%%%%%%%%%%%%%%%%%%%%%%%%%%%%%%%%%%%%%%%%%%%%%%%%%%
%%%%%%%%%%%%%%%%%%%%%%%%%%%%%%%%%%%%%%%%%%%%%%%%%%%%%%%%%%%%%%%%%%%%%

\section{DMFT Instability}

\begin{figure}
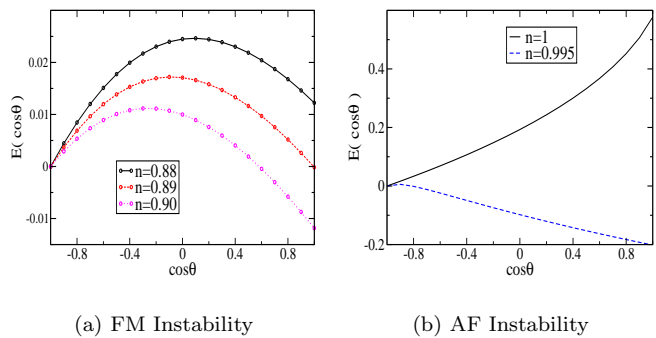

%\vspace*{1cm}
   \subfigure[FM Instability]{\epsfig{file=ESz_J8.eps,  height=1.4in,
   width=1.6in}}
   \hspace{0.2cm}
   \subfigure[AF Instability]{\epsfig{file=ESz_J1_AF.eps,  height=1.4in,
   width=1.6in}}
   \caption{(a) Effective potential for the core spin (Eqn(\ref{Eqs:Eeff})) 
   for a ferromagnetic state calculated at $T=0$, $J=8$, and densities shown. As we increase the 
   occupation number from 0.88 to 0.90, the minimum of $E_{eff}(\cos\theta)$ 
   changes from $\cos\theta=-1$ to 1 which indicates the FM DMFT solution
   becomes unstable when $n>0.89$. (b) Effective potential for the core spin 
   for an anti-ferromagnetic state calculated at $T=0$, $J=1$, and densities shown. 
   At $n=1$, the minimum of $E_{eff}$ happens
    at $\cos\theta=-1$, while at $n=0.995$ the minimum changes to $\cos\theta=1$.
    The AF DMFT solution only exists at $n=1$.  }
   \label{Fig:Instability}
\end{figure}

In this section we show that the DMFT equations
exhibit an apparently previously unnoticed instability. We begin
with $T=0$.
From Eqn(\ref{Eqs:DMFT1}),  the $G_{imp}$ is
\begin{eqnarray}
G_{imp,\uparrow(\downarrow)} = < \frac{a_{\downarrow(\uparrow)}-(+)J\cos\theta }
{a_{\uparrow}a_{\downarrow}-J \cos\theta
(a_{\uparrow}-a_{\downarrow})-J^2}>
\label{Eqs:DMFT3}
\end{eqnarray}
where $<...>$ means the angular average with respect to the weight
function $e^{-E_{eff}(\cos\theta)/T}$, with $E_{eff}$ defined in 
Eqn(\ref{Eqs:Eeff}) . At zero temperature, the
only contribution of the angular average is from the absolute
minimum of $E_{eff}(\cos\theta)$. To find the DMFT solution at
$T=0$, one first {\em assumes} the absolute minimum of
$E_{eff}(\cos\theta)$ occurs at a fixed value, for example
$\cos(\theta) =-1$, obtains
$G_{imp,\uparrow(\downarrow)}^{-1}=a_{\downarrow(\uparrow)}-(+)J$
from Eqn(\ref{Eqs:DMFT3}), and gets the self energy
$\Sigma_{\uparrow\,(\downarrow)}=+(-)J$ from Eqn(\ref{Eqs:DMFT1}).
Finally, one uses the $a$ obtained by the above procedure to
calculate $E_{eff}(\cos\theta)$ to see if the minimum is located
at the point originally assumed. Note that different ground states
(FM, AF..) enter the above procedure only via the self consistent
equation Eqn(\ref{Eqs:DMFT2-0}) (or Eqn(\ref{Eqs:DMFT2})).

Fig \ref{Fig:Instability} shows that as density is increased at
fixed large $J$, the self consistency breaks down, in an unusual
manner: $E_{eff}(\cos\theta)$ remains locally stable (slope
around the assumed minimum $\cos(\theta)=-1$ remains positive) but
the global minimum of $E_{eff}$ moves to $\cos(\theta)=1$. In the
regime where this phenomenon occurs, no solution of the DMFT
equations exists. Any initial solution we have considered leads to
a similar inconsistency (as is shown in panel b of Fig\ref{Fig:Instability} 
for the case of anitferromagnetism).

\begin{figure}[htbp]
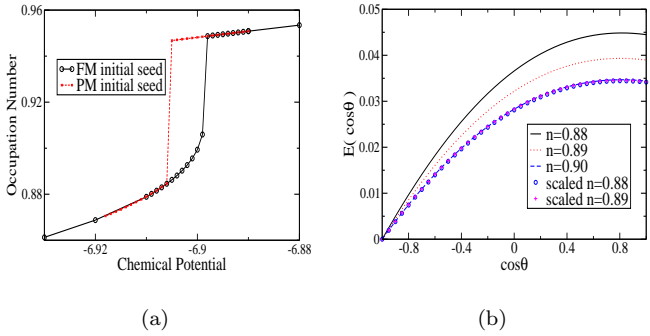

%\hspace*{0.8cm}
%\vspace{0.4cm}
  \subfigure[]{\epsfig{file=Nmu_J8_T0.01.eps, height=1.4in, width=1.6in}}
   \hspace{0.2cm}
   \subfigure[]{\epsfig{file=ESz_J8_T0.01_n-2.eps, height=1.4in, width=1.6in}}
   \caption{(a) $n(\mu)$ at $J=8$ and $T=0.01$, showing two phase behavior.
    Solid line: obtained from the initial seed close to the FM state. 
    Dashed line: obtained from the initial seed close to the PM state. 
   (b) $E_{eff} (\cos\theta)$ at $J=8$, $T=0.01$, $n=0.88$(solid line), 0.89
   (dot line), and 0.90(dashed line). The circle and dagger represent the scaled
   effective potential for $n=0.88$ and 0.89 which are almost indistinguishable to the $E_{eff}$
   at $n=0.90$.}
   \label{Fig:N(mu)T0.01}
\end{figure}

This instability is also manifest at $T>0$. As $T$ is decreased at fixed $n$, 
the convergence becomes slower and below some 
temperature $T^*(n)$, no stable solution can be found for a $J$-dependent range of $n$.
The absence of a solution for some range of $n$ can be seen in a different
way by solving the model as a function of chemical potential $\mu$
Fig \ref{Fig:N(mu)T0.01}(a) shows that as $\mu$ is increasedat fixed low $T$,
a first order transition occurs to a paramagnetic state, with a corresponding
jump in $n$. Associated with the first order transition is a coexisting region 
in which two solutions are locally stable (FM with lower $n$ and PM with
higher $n$); the DMFT equations correspondingly have two solutions, which one
is found depends on the initial seed.
The solid and dashed lines in Fig \ref{Fig:N(mu)T0.01}(a) are obtained
from initial seeds close to FM and PM states respectively.

The absence of convergence
may be understood from the density dependent effective potential,
shown e.g. in Fig \ref{Fig:N(mu)T0.01}(b). One sees that as $n$ is
increased, $E(\cos\theta=1)$ decreases; this is a precursor of the effect
shown in Fig \ref{Fig:Instability}(a). Indeed, the curve $E(\cos\theta)$
is reduced by an $n$-dependent scale factor. For $n$ larger than a critical value
(here $n\sim0.905$), $E(\cos\theta=1)$ is small enough relative to the temperature
that this region begins to contribute to $<\cos\theta>$, lowering the maximum $m$ 
that can be sustained and destabilizing the ferromagnetic solution.

%----------------------------------------------------------------------------------
\begin{figure}[htbp]
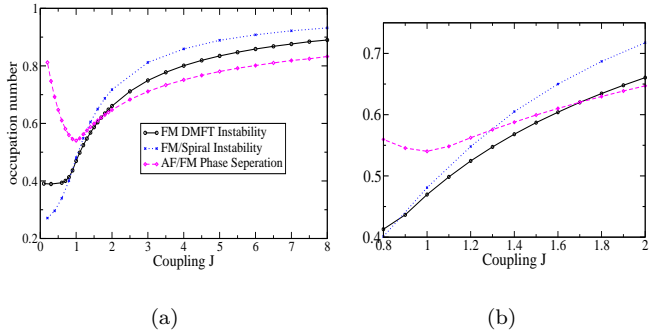

%\vspace{0.4 cm}
%\hspace*{1cm}
%\includegraphics[width=\hsize]{Fig111.eps}
   \subfigure[]{\epsfig{file=3PhaseBoundary_ncJ.eps, height=1.4in, width=1.7in}}
   \hspace{0.2cm}
   \subfigure[]{\epsfig{file=3PhaseBoundary_ncJ-2.eps, height=1.3in, width=1.5in}}
   \caption{(a) $T=0$ phase boundaries. Solid line: DMFT instability above which the 
   DMFT solution becomes unstable  
   Dot line: spiral instability line. The system is
   spiral(FM) above(below) this line. Dashed line: AF/FM phase separation. (b) 
   Expansion of the region $0.8t<J<2t$. For $n$, $J$ in the triangle bounded below
   by the solid line and above by the dashed and dot lines, the ground state is not known. }
   \label{Fig:PhaseDiagram}
\end{figure}
%--------------------

%----------------------------------------------------------------------------------
\begin{figure}[hbt]
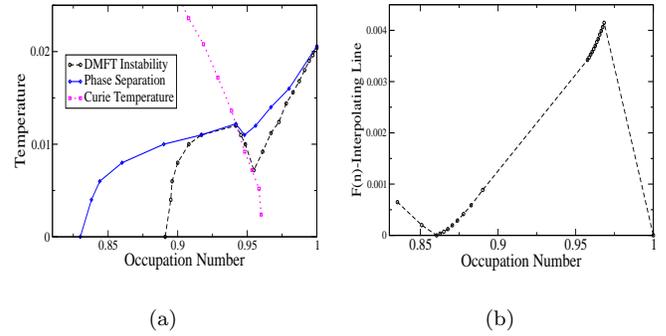

%   \centering
%\vspace{0.3 cm}
%\hspace*{-0.5cm}
%\includegraphics[width=\hsize]{Fig111.eps}
   \subfigure[]{\epsfig{file=J8_T_PhaseDiagram.eps, height=1.4in, width=1.6in}}
   \hspace{0.2cm}
   \subfigure[]{\epsfig{file=En-LinearOffset_J8_T0.008.eps, height=1.4in, width=1.6in}}
   \caption{ (a) Phase Boundaries for J=8 in $n-T$ plane. Solid line: the phase separation
   boundary obtained by Maxwell construction. Dotted line: Curie temperature. 
   Dashed line: the boundary of DMFT instability.
   (b) Maxwell construction for $J=8$, $T=0.008$ -- the 
   difference between free energy and interpolating line. The ground
   state is FM when $n<0.86$, and FM/AF phase separaion for $0.86<n<1$.  }
   \label{Fig:PST0.01}
\end{figure}
%--------------------

%%%%%%%%%%%%%%%%%%%%%%%%%%%%%%%%%%%%%%%%%%%%%%%%%%%%%%%%%%%%%%%%%%%%%%%%%%%%%
%%%%%%%%%%%%%%%%%%%%%%%%%%%%%%%%%%%%%%%%%%%%%%%%%%%%%%%%%%%%%%%%%%%%%%%%%%%%%%
\section{Interpretation}

We argue in this section that the DMFT instability documented in the previous section
is a manifestation of competing instabilities (primarily phase separation)
in the original model. To establish this we show in the panel (a) of Fig \ref{Fig:PhaseDiagram}
a $T=0$ phase diagram in the density-coupling plane.
The dash-dot line shows the phase separation boundary obtained from the global
energy computation; for $n$ above this line the model phase separates into
an $n=1$ AF and an $n<1$ FM state. The dotted line shows the phase boundary
between uniform FM and spiral states. Finally, the heavy solid line shows the
region above which the FM DMFT solution is unstable at $T=0$. When $J$ is large enough
that the FM state is fully polarized ($J>2t$), we see that the DMFT instability
line follows the phase separation line, but is inside the region of
phase separation. We therefore suggest that in this region the DMFT instability is a consequence of phase separation and this DMFT instability line can be used as a rough
estimate of the real phase separation boundary.

When $J<0.8 t$, the DMFT instability indicates the presence of a spiral 
state with lower energy
than the ferromagnetic state.
For $0.8t<J<1.7t$ (Fig \ref{Fig:PhaseDiagram}(b)), there exists a narrow 
region of $n$ where none of the uniform phases we considered solve the DMFT equations
and the Maxwell constructions seem not to indecate phase separation. 
We believe that in this region there exists a uniform non FM/AF/spiral/paramagnetic state
(either the ground state or the phase separation beteen FM and that state)
which we do not know yet.

At $T>0$ the situation is similar. The DMFT instability is contained inside the regime of
phase separation. For example, we show in Fig \ref{Fig:PST0.01}(a) the 
phase diagram and the range of DMFT instability
in the density-temperature plane for $J=8t$. The heavy line shows the boundary
of the regime of phase separation obtained by Maxwell construction: 
for $0<T<0.011$, the phase separation is 
between FM and AF($n=1$); for $0.012<T<0.02$, the phase separation is 
between PM and AF($n=1$); for $0.011<T<0.012$, the phase separation is
either PM-AF($n=1$) or FM-PM\cite{FM-PM} according to the location $n,T$ relative to 
the homogenous Curie temperature (dotted line). 
The dashed line shows the region where the DMFT solution fails to converge at
that given density $n$ (the DMFT equation has stable solution for all $\mu$,
see Fig \ref{Fig:N(mu)T0.01}(a)). For $n>0.95$, the DMFT instability line denotes the 
temperature below which (a) the paramagnetic state is linearly unstable to
antiferromagnetic and (b) no stable antiferromagnetic solution exists (except $n=1$).

Fig \ref{Fig:PST0.01}(b) shows the results of a Maxwell construction for $J=8$ and
that at $T=0.008$, presented as the difference between calculated free energy $F(n)$
and the interpolating line $I(n) = F(n=1)+\frac{F(n^*)-F(n=1)}{1-n^*}(1-n)$ with
$n^*=0.86$. Phase separation is seen to occur for $n^*<n<1$, while the DMFT
instability range is $0.895<n<0.95$ and $0.96<n<1$.

\section{Conclusion}

We have found an instability in the ferromagnetic DMFT equation
for the single site double exchange model and shown that this
instability corresponds to the FM/AF phase separation when the
coupling $J$ is larger than half bandwidth (2$t$) and to another
ground state (spiral) in the small coupling region. There exists a
small window, around intermediate $J$, where no stable FM DMFT
solutions exist while the spiral or phase separation is not the
ground state, and we believe there is a non FM/AF/Spiral/Para
ground state existing in this region.  We have presented evidence
that the instability is a signal, obtained from a calculation at a
fixed parameter value, of the existence of an instability
(typically phase separation) which normally is established via a
global computation, comparing free energies at many different
parameter values. We therefore propose that the DMFT instability
is a computationally convenient way to estimate the boundary of
phase separation.

%%%%%%%%%%%%%%%%%%%%%%%%%%%%%%%%%%%%%%%%%%%%%%%%%%%%%%%%%%%%%%%%%%%%%%%%%%%%%%%%%%%
%\section*{Acknowledgements}
We thank Dr. Satoshi Okamoto for many helpful discussions.
This work is supported by DOE ER46169 and Columbia University
MRSEC.

%%%%%%%%%%%%%%%%%%%%%%%%%%%%%%%%%%%%%%%%%%%%%%%%%%%%%%%%%%%%%%%%%%%%%%%%%%%%%%%%%%%%%%%%%%%%%%%%

%%%%%%%%%%%%%%%%%%%%%%%%%%%%%%%%%%%%%%%%%%%%%%%%%%%%%%%%%%%%%%%%%%%%%%%%%%%%%%%%%%%%%%%%%%%%%%%%
%%%%%%%%%%%%%%%%%%%%%%%%%%%%%%%%%%%%%%%%%%%%%%%%%%%%%%%%%%%%%%%%%%%%%%%%%%%%%%%%%%%%%%%%%%%%%%%%


\begin{thebibliography}{10}

\bibitem{Georges}
A.Georges, G.Kotliar, W.Krauth, and M.Rozenberg, Rev of
Modern Physics {\bf 68}, 13 (1996)

\bibitem{Hubbard}
A.Georges, and W. Krauth, Phys.Rev.Lett {\bf 69}, 1240  (1992).
M.J. Rozenberg, G.Kotliar, amd X.Y. Zhang, Phys.Rev.Lett
{\bf 69}, 1236  (1992).


\bibitem{Kondo}
A. J. Millis, R. Mueller, and Boris I. Shraiman 
Phys. Rev. B {\bf 54}, 5389 (1996). 
A. J. Millis, R. Mueller, and Boris I. Shraiman 
Phys. Rev. B {\bf 54}, 5405 (1996).


\bibitem{Chattopadhyay}
A.Chattopadhyay, A.J.Millis, and S. Das Sarma
Phys.Rev.B. {\bf 64}, 012416-1 (2001)

\bibitem{Dagotto2}
E.Dagotto Sience {\bf 309}, 257 (2005)

\bibitem{Colossal}
See, e.g. the articles in {\em Colossal Magnetoresistive Oxides},
Y.Tokura,ed (Gordon and Breach: Tokyo, 1999)

\bibitem{FE}
In Eqs(\ref{Eqs:LatFreeEnergy}), for FM and AF states, the lattice Green function
$ G(i\omega, \vec{k}) $ 
are indicated in Eqn(\ref{Eqs:DMFT2-0}) and
Eqn(\ref{Eqs:DMFT2}), while the trace are over
full and reduced Brillouin zone respectively.


\bibitem{Spiral}
J.Inoue and S.Maekawa
Phys.Rev.Lett {\bf 74}, 3407  (1995).

\bibitem{Dagotto}
S.Yunoki, J.Hu, A.L.Malvezzi, A.Moreo,
N.Furukawa, and E.Dagotto
Phys.Rev.Lett {\bf 80}, 845  (1998).

\bibitem{Guinea}
D.P.Arovas, G.Gomez-Santos, F.Guinea, 
Phys.Rev.B. {\bf 59}, 13569 (1999)

\bibitem{FM-PM}
The precise boundary of PM-FM phase separation is hard to determine because
it is very close to the FM DMFT instability line. They coincide within our numerical error.



%\bibitem{h-unit}
%For converting the magnetic field from energy unit (ev) into normal unit (gauss or tesla),
%we use the following identity
%\begin{eqnarray}
%hm \rightarrow h\left[ \frac{7}{2}g\mu_e\right]m\sim h (7\mu_e)m
%\end{eqnarray}
%where $\mu_e$ is the Bohr magneton $\mu_e=5.79\times 10^{-9}$ev/gauss.
%On the left side, $h$ has unit of energy while on the right hand side, $h$ has
%normal magnetic field unit. We approximate $g=2$ in the above equation since the spin-orbit
%coupling is small for Europium. So in this case, $1$ ev = $2.467 \times 10^4$ tesla.

\end{thebibliography}
\end{document}